\documentclass[prb,twocolumn,floatfix,notitlepage,superscriptaddress,longtable]{revtex4-2}
\usepackage{amsmath}
\usepackage{lipsum} 
\usepackage{verbatim}
\usepackage{epsfig}
\usepackage{subfigure}
\usepackage{graphicx}
\usepackage{amsfonts}
\usepackage[figuresright]{rotating}
\usepackage{amssymb}
\usepackage{amsmath}
\usepackage[normalem]{ulem}
\usepackage{psfrag}
\usepackage{esint} 
\usepackage{bm}
\usepackage[colorlinks,linkcolor=blue,anchorcolor=blue,citecolor=blue,urlcolor=blue]{hyperref}
\usepackage[version=4]{mhchem}
\usepackage[svgnames]{xcolor}

\def\be{\begin{equation}} \def\ee{\end{equation}}
\def\bea{\begin{eqnarray}} \def\eea{\end{eqnarray}}

\renewcommand{\vec}[1]{\mathbf{#1}}

\newcommand{\ket}[1]{| #1 \rangle}

\newcommand{\braket}[2]{\langle #1 |#2\rangle}

\def\bpm{\begin{pmatrix}} \def\epm{\end{pmatrix}}
\usepackage{graphicx} 
\usepackage{xcolor}   
\usepackage{booktabs}
\usepackage{tikz}
\usetikzlibrary{positioning, shapes.geometric, arrows.meta, calc, fit, shadows}
\usetikzlibrary{arrows.meta} 
\usepackage{pgfplots}
\pgfplotsset{compat=1.18}    

\usepackage{tikz}
\usepackage{graphicx} 
\usetikzlibrary{positioning, shapes.geometric, arrows.meta, calc, fit, shadows, backgrounds}

\definecolor{boxblue}{RGB}{218, 232, 252}
\definecolor{boxgreen}{RGB}{213, 232, 212}
\definecolor{boxorange}{RGB}{255, 204, 153}
\definecolor{boxpurple}{RGB}{225, 213, 231}
\definecolor{addyellow}{RGB}{255, 255, 153}
\definecolor{arrowblue}{RGB}{22, 79, 134}
\definecolor{arrowred}{RGB}{178, 34, 34}
\definecolor{panelgray}{RGB}{245, 245, 245}

\newcommand{\UU}{\hat U}
\newcommand{\HH}{\hat H}
\newcommand{\sig}{\sigma}

\newcommand{\abs}[1]{\left|#1\right|}

\newcommand{\edited}[1]{{\color{Red}}}
\definecolor{Qicolor}{RGB}{3, 136, 252}

\makeatletter
\newcommand*{\balancecolsandclearpage}{%
  \close@column@grid
  \clearpage
}
\makeatother

\begin{document}
\title{Universal Neural Propagator: Learning Time Evolution in Many-Body Quantum Systems}

\author{Zihao Qi}
\email[Contact author: ]{zq73@cornell.edu}
\affiliation{Department of Physics, Cornell University, Ithaca, NY 14853, USA.}

\author{Christopher Earls}
\affiliation{Center for Applied Mathematics, Cornell University, Ithaca, NY 14853, USA.}

\author{Yang Peng}
\email[Contact author: ]{yang.peng@csun.edu}
\affiliation{Department of Physics and Astronomy, California State University, Northridge, Northridge, CA 91330, USA}
\affiliation{Institute of Quantum Information and Matter and Department of Physics,California Institute of Technology, Pasadena, CA 91125, USA}

\date{\today}

\begin{abstract}
Conventional approaches to simulating quantum many-body dynamics produce a single trajectory: if the Hamiltonian or the initial state is changed, the computation must be re-performed. Recent efforts toward foundation models have begun to address this limitation, yet existing methods transfer across either Hamiltonians or initial states, but not both. In this work, we introduce the Universal Neural Propagator (UNP), a single, unified model that learns the functional mapping from driving protocols to time-evolution propagators. Trained in an entirely self-supervised way, a single UNP model predicts dynamics across a function space of driving protocols and an exponentially large Hilbert space of initial states simultaneously. We benchmark on a two-dimensional driven Ising model and demonstrate the UNP's accuracy and transferability across product and entangled initial states, as well as for both in- and out-of-distribution driving protocols. The UNP remains accurate at system sizes beyond exact diagonalization, and can be efficiently fine-tuned across all initial states using observable data. By shifting the object of learning from quantum states to operators, this work opens a route toward transferable simulation of driven quantum matter.

\end{abstract}

\maketitle

\section{Introduction}
Simulating the real-time dynamics of quantum many-body systems remains one of the central challenges in computational physics. This difficulty arises not only due to the exponential scaling of the Hilbert space dimension with system size, but also from the growth of entanglement under time evolution. Tensor-network methods provide powerful alternatives to exact numerical methods in one dimension and for weakly entangled regimes, but their efficiency can be limited in higher dimensions and non-equilibrium settings~\cite{dmrg_review1, dmrg_review2, dmrg1, dmrg2, mps_dmrg, iPEPS, PEPS}.

\emph{Neural quantum states (NQS)} have recently emerged as a promising approach for representing many-body wavefunctions using neural networks. By parameterizing the complex amplitudes of a quantum state, NQS leverage the expressivity of neural network architectures and efficiently capture correlations across the full Hilbert space. NQS approaches have been remarkably successful for problems such as ground-state search~\cite{carleo2017, autoregressive-hubbard, Transformer, NN_GS, NN_GS2, NN_GS3} and real-time evolution~\cite{tdvp, NQS_dynamics_tDVP, tNQS_Bohrdt, tNQS_Carleo, real-time-evolution-nqs}. 

However, most approaches to simulating quantum dynamics, whether based on tensor networks or neural quantum states, remain tied to a specific configuration: for a given Hamiltonian and initial state, one optimizes parameters in a variational representation of the time-dependent wavefunction. If either the Hamiltonian or the initial state changes, the
entire computation must generally be repeated from scratch.

Recent works have started to move beyond this instance-by-instance paradigm.
Inspired by the success of foundation models in machine learning~\cite{foundation_model}, several works have proposed neural networks conditioned on Hamiltonian parameters,
producing a single model that represents ground states across a family of
Hamiltonians~\cite{Transformer, rende2025foundation,
zaklama_foundation_attention, zaklama2026large}. In the dynamical setting,
neural-operator-based models have extended this idea by learning to evolve
states across families of time-dependent driving
protocols~\cite{NOQS}. In a complementary direction, efforts have instead targeted
transferability across initial states, learning representations of the
quantum propagator under a fixed
Hamiltonian~\cite{FNOFloquet, Shah2026FNOQuantumSpin,
Zhang2025NeuralQuantumPropagators}.

In all prior works, however, only ``half of the problem'' has been addressed: existing models \textit{either} transfer across
Hamiltonians for a fixed initial state, \textit{or} across initial states for a fixed Hamiltonian. 
There is, then, a need for a more universal propagator: a single model that transfers across a wide range of Hamiltonians, as well as initial states. Such a model is particularly desirable for applications that require
efficiently scanning the joint space of driving protocols and initial
states, such as quantum optimal control, where one seeks a driving
sequence that steers a system toward a target state and must verify robustness across initial preparations~\cite{optimal_control_1, optimal_control_2, optimal_control_3}.

In the current work, we bridge this gap by introducing the \emph{Universal Neural Propagator (UNP)}, a single, unified model that learns the many-body time-evolution operator directly, transferring across both an infinite-dimensional function space of driving protocols and an exponentially large Hilbert space of initial states simultaneously. The key idea of the UNP is to learn a \emph{functional} mapping from driving protocols $H(t)$ to time-evolution propagators $U(t)$, which are represented as neural quantum states in a doubled Hilbert space. Once trained, for any unseen driving fields, the UNP predicts the corresponding propagator. The model is trained in an entirely self-supervised fashion, requiring no
pre-computation or external data.

We validate the UNP on a two-dimensional driven transverse-field Ising
model, demonstrating accurate reproduction of exact time evolution across
multiple product and entangled initial states, as well as both in- and out-of-distribution driving
protocols. We further show that the pre-trained propagator can be
efficiently fine-tuned using observable data, thereby improving accuracy across all initial states. We also demonstrate the UNP model is scalable and remains accurate for system sizes beyond the reach of exact methods.

These results establish the UNP as a concrete step toward transferable neural simulation of many-body quantum dynamics. Unlike conventional
approaches that must be re-run for each combination of Hamiltonian
and initial state, a single trained UNP can be reused across protocols and
initial states with no additional training or optimization required. More broadly, the UNP demonstrates
that neural architectures can represent not just individual quantum states,
but the operators that govern time-evolution. By shifting the learning target from
states to propagators, our framework provides a new paradigm toward foundation models for quantum dynamics.

The remainder of the paper is organized as follows. In Sec.~\ref{sec:review}, we review background information on neural quantum states. In Sec.~\ref{sec:UNP}, we introduce the Universal Neural Propagator framework: we describe the model's hybrid FNO–transformer architecture and its self-supervised training procedure. In Sec.~\ref{sec:results}, we present numerical results on the two-dimensional driven transverse-field Ising model and demonstrate the model's excellent accuracy and transferability. We conclude with a discussion of the results and outlook in Sec.~\ref{sec:discussion}.

\section{Background \label{sec:review}}
In this section, we review the background information on Neural Quantum States (NQS), focusing on autoregressive representations of quantum states, estimation of expectation values, and recent progress toward NQS foundation models.
 
For concreteness, we focus on systems consisting of $N$ spin-$1/2$ degrees of freedom. Our formalism extends naturally to systems with higher spins.
A quantum state $|\psi\rangle$ is fully specified by its amplitudes
in a computational basis $\{|\alpha\rangle\}$:
\begin{equation}
  \psi(\alpha) = \langle \alpha|\psi\rangle.
  \label{eq:wfn}
\end{equation}

In this work, we take the computational basis to be the $z$-basis, $\alpha=(\alpha_1,\dots,\alpha_N)\in\{0,1\}^N$. Quantum states, therefore, can be viewed as functions that map from the configuration space (in this case, bit-strings of length $N$) to complex amplitudes:
\begin{equation}
    \psi: \alpha \rightarrow \psi(\alpha), \qquad \psi(\alpha) \in \mathbb{C}.
\end{equation}

Neural quantum states (NQS) parameterize wavefunctions with neural networks, which are universal functional approximators~\cite{NNunivapprox1, NNunivapprox2, NNunivapprox3}. Specifically, NQS aims to output the complex amplitude
\begin{equation}
  \psi_\theta({\alpha})
  = \exp\left( \frac{1}{2}\log p_\theta({\alpha})
        \,+\, i\,\phi_\theta({\alpha})\right),
  \label{eq:nqs}
\end{equation}
for any configuration $\alpha$. Here $\theta$ denotes the set of parameters in the NQS, $p_\theta({\alpha})=|\psi_\theta({\alpha})|^2$ is the Born probability, and $\phi_\theta({\alpha})$ is the
complex phase.

Various network architectures have been used to represent quantum states,
including restricted Boltzmann machines~\cite{RBM_review, RBM_symmetry, RBM_symmetry2, carleo2017}, convolutional networks~\cite{CNN1, CNN2, tdvp, CNN4},
and recurrent networks~\cite{RNN, RNNauto, RNN2, sharir2020_autoregressive, RNN4, RNN5}. Among these architectures, \textit{autoregressive} NQS have been remarkably successful due to the ability to factorize the probability:
\begin{equation}
  p_\theta(\alpha)
  = \prod_{i=1}^{N}
    p_\theta(\alpha_i\,|\,\alpha_1,\dots,\alpha_{i-1}),
  \label{eq:ar_prob}
\end{equation}
with an analogous factorization of the phase,
$\phi_\theta(\alpha)$. Here $p(\cdot | \cdot)$ denotes the conditional probability. One direct consequence is exact, independent sampling for autoregressive models, thereby producing uncorrelated samples.
This circumvents the autocorrelation overhead inherent
in Markov-chain Monte Carlo (MCMC) and avoids the need to normalize over exponentially large configuration spaces~\cite{nqs_review1}.

With samples drawn from the probability distribution, one can estimate expectation values of observables with respect
to $\ket{\psi_\theta}$. Expectation values can be written as an average over the
Born distribution $p_\theta(\alpha)$:
\begin{equation}
  \langle\hat{O}\rangle
  = \sum_{\alpha} p_\theta({\alpha})\,O_{\mathrm{loc}}({\alpha}),
\end{equation}
where
\begin{equation}
    O_{\mathrm{loc}}({\alpha})
  = \sum_{{\alpha}'}
    \frac{O_{{\alpha}{\alpha}'}\,\psi_\theta({\alpha}')}
         {\psi_\theta({\alpha})}
  \label{eq:local_est}
\end{equation}
is the local estimator of the operator $\hat{O}$. The expectation value is then approximated by a sample mean,
\begin{equation}
  \langle\hat{O}\rangle
  \approx \frac{1}{M}\sum_{m=1}^{M}
    O_{\mathrm{loc}}\!\bigl({\alpha}_{m}\bigr),
  \label{eq:mc_avg}
\end{equation}
with samples ${\alpha}_{m}\sim p_\theta({\alpha})$ drawn from the probability distribution. For spatially localized operators (such as energy of local Hamiltonians), each evaluation of $O_{\mathrm{loc}}$
involves only polynomially many nonzero matrix elements and can be performed efficiently.

The NQS framework has been remarkably successful in various problems in quantum many-body systems. For example, finding the ground state of a system amounts to optimizing the parameters $\theta$ to minimize the energy, $E = \langle \hat{H} \rangle$~\cite{carleo2017, RBM_symmetry, sharir2020_autoregressive, RNNauto}. Quantum dynamics can also be simulated within this framework through the time-dependent variational principle
(TDVP).
Given a Hamiltonian $H(t)$, one projects the exact time evolution onto the variational manifold by minimizing the distance between $\partial_t \ket{\psi_\theta(t)}$ and $-iH \ket{\psi_\theta(t)}$ at each time, yielding equations of motion for the parameters $\theta$. TDVP-based NQS dynamics has been applied to quench problems
in one and two dimensions, and has demonstrated competitive performance against traditional numerical methods~\cite{carleo2017, tdvp, tNQS_Bohrdt, tNQS_Carleo, real-time-evolution-nqs, real-time-evolution-nqs}. However, despite the expressive ansatz, the variational trajectory is not transferable: changing either the initial state or the Hamiltonian $H(t)$ generally requires a new variational computation from scratch.

Several recent works have begun to close this gap, through the development of ``foundation models'': networks conditioned on Hamiltonian parameters,
producing a single model that represents ground states
across a family of Hamiltonians~\cite{rende2025foundation, Transformer, zaklama_foundation_attention, zaklama2026large}. However, the efforts thus far have mostly targeted equilibrium
properties. Ref.~\cite{NOQS} makes a concrete step toward a foundation model for quantum dynamics, generalizing across a functional space of driving protocols, but the approach still optimizes for a fixed initial state. In Sec.~\ref{sec:UNP}, we introduce the Universal Neural Propagator (UNP), a single model that transfers across driving protocols \textit{and} initial states simultaneously.

\section{Universal Neural Propagator \label{sec:UNP}}
Under time evolution governed by a Hamiltonian $\hat H(t)$, quantum states evolve according to Schr\"odinger's equation:
\begin{equation}
    i \partial_t \ket{\psi(t)} = \hat H(t) \ket{\psi (t)}.
\end{equation}

Formally, the time-evolved quantum state can be written as $\ket{\psi(t)} = \UU(t) \ket{\psi_0}$,
where $\ket{\psi_0}$ is the initial state, and the propagator is defined as
\begin{equation}
    \UU(t) = \mathcal{T} \exp\left({-i\int_0^t \hat H(t') dt'} \right),
    \label{eq:propagator-definition}
\end{equation}
with $\mathcal{T}$ being the time-ordering operator. Note that $\UU(t)$ explicitly depends on the Hamiltonian $\hat H(t)$.

\begin{figure*}
    \centering
\includegraphics[width=0.8\linewidth]{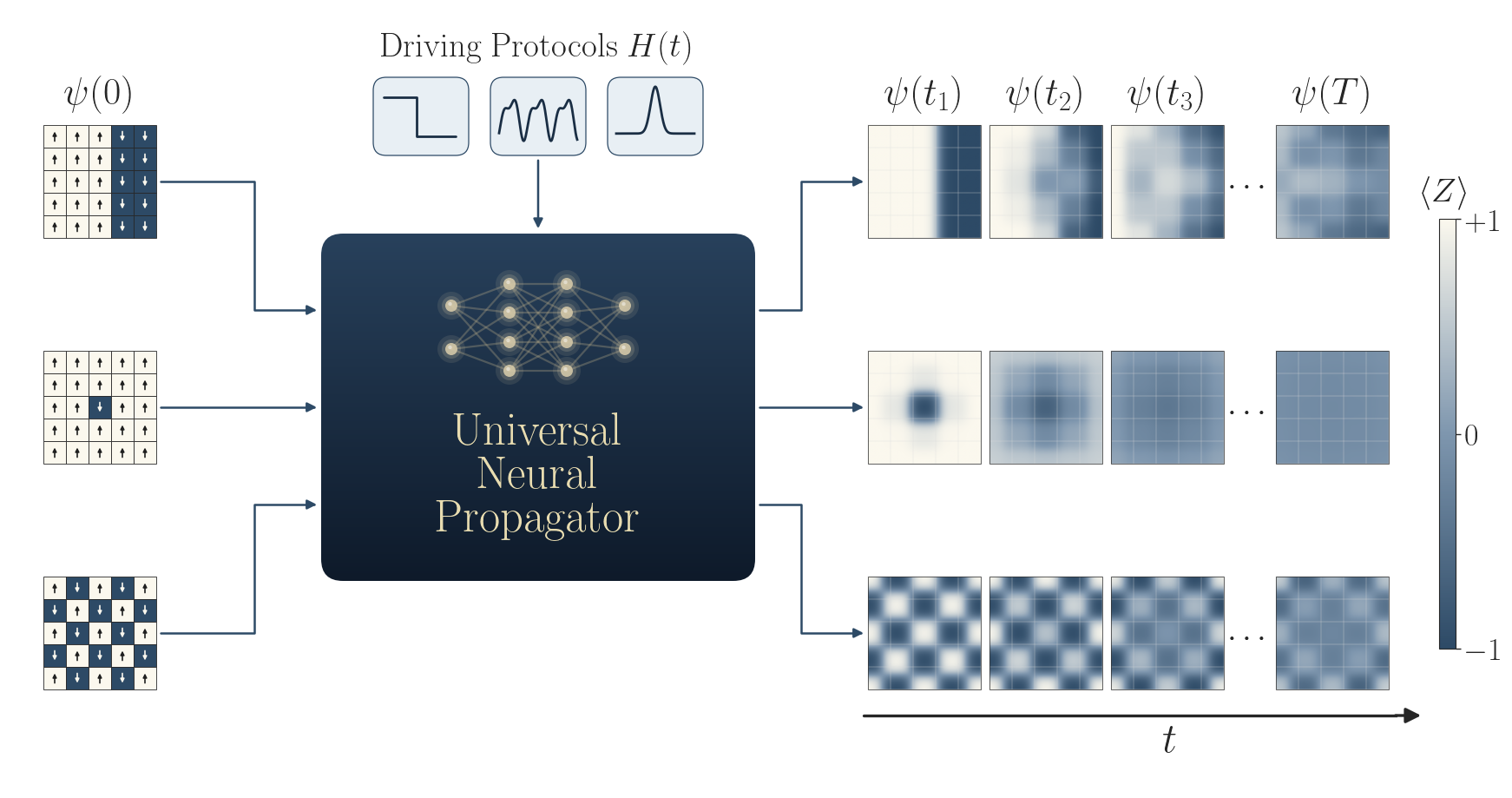}
    \caption{Illustration of the Universal Neural Propagator (UNP) framework. For any driving protocol $\hat H(t)$, the corresponding time-evolution quantum propagator $\UU(t)$ is predicted by a neural-network-operator hybrid UNP architecture. The learned propagator can be further applied to track time evolution starting from arbitrary initial states in a given computational basis. A single, unified UNP model allows for transferring across an infinite-dimensional functional space of driving protocols \textit{and} an exponentially large space of initial states, predicting time dependent observables at any time in the training interval.}
    \label{fig:illustration}
\end{figure*}

In this work, instead of tracking time evolution at the level of quantum states, we aim to directly learn to predict the time-evolution \textit{propagator} $\UU(t)$, using driving protocols $\hat H(t)$ as inputs. At a high level, we aim to form a neural-operator-based representation of time evolution itself: once trained, the Universal Neural Propagator (UNP) can be used to efficiently evaluate $\UU(t)$, transferring over a functional space of driving protocols. That is, the UNP model can be viewed as learning a \textit{functional} mapping:
\begin{equation}
    \text{UNP}: \hat H(t) \longrightarrow \widetilde{U}(t),
\end{equation}
where $\widetilde{U}(t)$ denotes the UNP's predicted time-evolution operator. Using $\widetilde U (t)$, we can \textit{further} transfer across different initial states in a computational basis, and track their time evolution under the given Hamiltonian $\hat H(t)$. General initial states, which can be expressed as linear combinations of computational-basis states, can in principle be evolved by linearity. The concept of the UNP framework is illustrated in Fig.~\ref{fig:illustration}. In the following sections, we describe the doubled-space representation of $\UU(t)$, the UNP architecture, and the training procedure.

\subsection{Time Evolution Propagator in Doubled Space}
From its definition, Eq.~\ref{eq:propagator-definition}, the propagator $\UU(t)$ satisfies
\begin{equation}
i\partial_t \UU(t) = \HH(t)\UU(t),
\end{equation}
with the initial condition $\UU(0)=I.$ In a computational basis $\{ \ket{\alpha} \}$ (fixed to be the $z$-basis throughout), matrix elements of $\UU(t)$ satisfy
\begin{equation}
i\partial_t U_{\alpha\beta}(t)
=
\sum_{\alpha'}
H_{\alpha\alpha'}(t)\,
U_{\alpha'\beta}(t),
\label{eq:column_schrodinger}
\end{equation}
where $U$ and $H$ are the matrix representations of the operator $\UU$ and $\hat H$ in this basis, respectively.

We can normalize the propagator through dividing by $\sqrt{D}$, where $D=2^N$ is the Hilbert space dimension, so that the full doubled-space norm is
\begin{equation}
\sum_{\alpha,\beta}\abs{ U_{\alpha\beta}(t)}^2 = 1.
\label{eq:frob_norm}
\end{equation}

In what follows, $U(t)$ denotes the \textit{normalized} propagator. Given a fixed initial computational-basis state $\ket{\beta}$, the corresponding evolved wavefunction is obtained by fixing the input index $\beta$ and reading off the column of the propagator,
\begin{equation}
    \ket{\psi_\beta(t)}_\alpha = U_{\alpha\beta}(t).
    \label{eq:psi_from_U_column}
\end{equation}
After the appropriate wavefunction normalization, this gives the time-evolved state starting from $\ket{\beta}$. Since the predicted propagator $\widetilde U(t)$ represents all input-output amplitudes, the same model can be queried for different initial basis states, and general initial states in this basis can then be evolved by linear superposition.

This motivates replacing the explicit matrix representation of the evolution operator by a neural representation of a single normalized state on a \textit{doubled} Hilbert space. In this representation, one copy of the Hilbert space labels the output configuration, while the second copy labels the input configuration. The construction is analogous to the vectorization of operators into states, as in the Choi--Jamiołkowski formalism~\cite{choi,jamiolkowski}.

For each site $i$, the pair $(\alpha_i,\beta_i)$ is encoded by a four-valued local token $\sigma_i \in \{0,1,2,3\}$, with the identification
\begin{align}
 \sigma_i=0 \leftrightarrow (\alpha_i, \beta_i) =(0,0);\quad 
 \sigma_i =1 \leftrightarrow (\alpha_i, \beta_i) =(0,1); \nonumber \\
 \sigma_i=2\leftrightarrow (\alpha_i, \beta_i) =(1,0);\quad
\sigma_i = 3\leftrightarrow (\alpha_i, \beta_i) = (1,1).
\end{align}

Thus, a doubled configuration
$\sig = (\sigma_1,\dots,\sigma_N)$
is equivalent to a \textit{pair} of basis strings $(\alpha,\beta)$. Conversely, a particular matrix element is recovered from doubled-space configuration through the following identification
\begin{equation}
(\sigma_1, \sigma_2, \dots, \sigma_N)
\leftrightarrow
\bigl( 2\alpha_1+\beta_1, 2\alpha_2 + \beta_2,\dots,2\alpha_N+\beta_N \bigr).
\end{equation}
Then
\begin{equation}
 U_{\alpha\beta}(t)
\leftrightarrow
U(\sig(\alpha,\beta);t),
\end{equation}
and we have mapped $\hat U(t)$ to a normalized \textit{state} $U(\sigma; t)$ that lives in doubled space. While the full propagator is nominally an exponentially large object, this representation allows for exploiting the NQS machinery: for local Hamiltonians, local estimators and variational gradients can be evaluated from a polynomial number of Monte Carlo samples, avoiding explicit summation over the exponentially large Hilbert space \cite{carleo2017, nqs_review1, nqs_review2}.

\subsection{Model Architecture \label{sec:architecture}}
\begin{figure}[t]
    \centering
\includegraphics[width=0.45\textwidth]{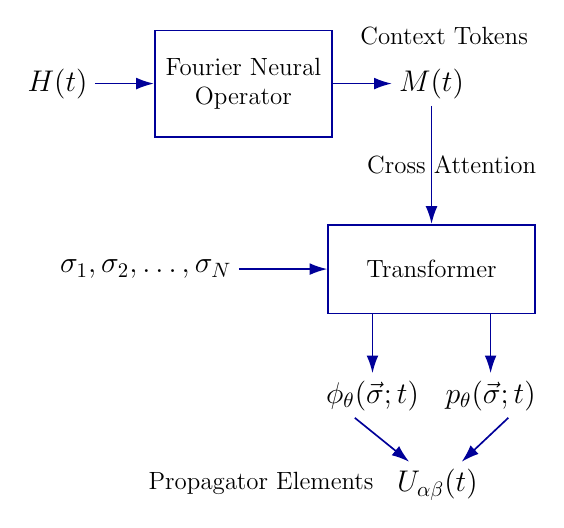}
    \caption{The architecture diagram of the UNP model. Driving protocols $H(t)$ are processed by a Fourier Neural Operator (FNO) and distilled into a set of context tokens $M(t)$ that encode the temporal structure of driving. The Transformer backbone of UNP processes this information through cross attention to assign an amplitude $p(\sigma; t)$ and phase $\phi(\sigma; t)$ for each spin configuration $\sigma$ in the doubled space. The phase and amplitude are combined to produce the matrix element $U_{\alpha \beta}(t)$.}
    \label{fig:architecture}
\end{figure}

To predict the normalized propagator, we employ a hybrid architecture that consists of a neural operator and an autoregressive transformer. At the highest level, the two components handle the temporal and spatial information respectively, and are coupled via the cross-attention mechanism. The architecture of UNP is illustrated in Fig.~\ref{fig:architecture}; in the following, we will discuss the role of each component. Further details about the model can be found in Appendix~\ref{app:architecture_details}.

We begin with the backbone of the UNP: a decoder-only transformer. As discussed in Sec.~\ref{sec:review}, the transformer architecture allows for autoregressive sampling over spin configurations and avoids the need to perform MCMC. The inputs to the transformer are configurations in the doubled space, represented by tokens $\vec{\sigma} = (\sigma_1,\dots,\sigma_N)$. The transformer model outputs the corresponding complex amplitude, $\widetilde U(\sig;t)$, which can be parameterized as:
\begin{align}
\widetilde U(\sig;t)
=
\exp\Bigl[
\frac12 &\sum_{i=1}^N \log p(\sigma_i | \sigma_{1},\dots,\sigma_{i-1}; t) \nonumber \\
&+ i\sum_{i=1}^N \phi(\sigma_i | \sigma_1, \dots, \sigma_{i-1}; t)
\Bigr],
\label{eq:autoregressive_logu}
\end{align}
where $p(\cdot |\cdot)$ and $\phi(\cdot | \cdot)$ parameterize the conditional amplitude and phase in the doubled space, respectively.

The transformer architecture consists of three components. An embedding layer first maps each doubled-space token $\sigma_i$ into a higher-dimensional latent space. Positional encoding is added to keep track of each site's spatial location. Following the embedding layer is a sequence of decoder blocks. Each decoder block captures the correlation among tokens via self-attention. The decoders also learn the propagator's dependence on driving protocols through cross-attention. Finally, the latent-space representations are projected back to the (conditional) amplitude and phases via an unembedding layer. We discuss the details of the transformer architecture, such as positional encoding, attention mechanism, and trainable parameters in Appendix~\ref{app:architecture_details}.

We next describe how the time-dependent driving protocols are encoded before being passed to the transformer. The protocol dependence enters the decoder only through the context tokens $M(t)$, which are generated by a Fourier Neural Operator (FNO). This choice is motivated by the fact that the input is not a fixed-dimensional parameter vector, but a function of time. Neural operators are precisely designed for this setting: rather than learning maps between finite-dimensional spaces, they learn mappings between function spaces~\cite{NeuralOperator_general,NeuralOperator2}. In particular, the FNO architecture is a natural choice, because it parameterizes the operator through learnable convolution in the frequency domain~\cite{FNO_Li}. FNO-based models have been successfully applied across a range of scientific problems, including many-body quantum systems~\cite{Shah2026FNOQuantumSpin,FNOFloquet,Zhang2025NeuralQuantumPropagators,wang2025fourierneuraloperatorapproach}.

Concretely, the driving protocol $H(t)$ is first lifted into a higher-dimensional latent representation. Each FNO layer then transforms this temporal representation to the frequency domain, applies a learnable spectral convolution to the frequency modes, and maps the result back to the time domain through an inverse Fourier transform. A final projection layer produces the context trajectory used by the transformer decoder. Because the spectral convolution couples information globally across the temporal domain, the FNO is well suited for compressing the full driving history into a compact representation. Importantly, the context tokens at different times are not generated independently. Instead, the neural operator processes the entire protocol as a function and outputs a correlated context trajectory. The FNO workflow is illustrated in Fig.~\ref{fig:architectural_details}(a), with additional architectural details provided in Appendix~\ref{app:architecture_details}.

The output of FNO is designed to be a \emph{velocity}
trajectory of the context tokens,
\begin{equation}
  \dot{M}(t) = \text{FNO}\!\bigl[H(t)\bigr].
  \label{eq:fno}
\end{equation}

The context tokens $M(t)$ can be obtained from integration of the predicted velocity,
\begin{equation}
  M(t) = M_0 + \int_0^t \dot{M}(\tau)\, d\tau,
  \label{eq:token_integration}
\end{equation}
where $M(0) = M_0$ is a fixed initial representation. We choose the FNO to learn the velocity field $\dot M(t)$ instead of the context tokens $M(t)$ by design; this choice is motivated by numerical stability. As we will discuss in Sec.~\ref{sec:training}, the loss function involves $\partial_t U$: using $\dot M(t)$ as FNO's output therefore avoids the need to perform numerical differentiation when the loss function is evaluated.

The transformer and FNO components are coupled through a \textit{cross}-attention mechanism. Intuitively, the context tokens $M(t)$, which encode information about the driving protocols, serve as values and keys. On the other hand, the transformer sends queries, and the output amplitudes and phases are \textit{conditioned} on the driving fields. This coupling enables generalization across driving protocols, as \textit{each} site's representation attends to the global driving history, through the learned context trajectory.

In summary, the UNP represents the time-evolution propagator as a normalized state in doubled-space. The time dependence is compressed into a small number of context tokens, generated from the driving protocol by an FNO. A transformer-based backbone learns the autoregressive probability distribution in the doubled space and outputs the propagator matrix elements at all times, $\widetilde U_{\alpha \beta}(t)$.

\subsection{Training Objective \label{sec:training}}
We now describe how the UNP model is trained. The training objective is built directly on the operator Schr\"odinger
equation, Eq.~\ref{eq:column_schrodinger}. Dividing Eq.~\ref{eq:column_schrodinger} through by $U_{\alpha\beta}(t)$ gives the
log-derivative form
\begin{equation}
  i \partial_t \log {U}_{\alpha\beta}(t)
  = \frac{\sum_{\alpha'} H_{\alpha\alpha'}(t)\, {U}_{\alpha'\beta}(t)}{{U}_{\alpha\beta}(t)}
  \;\equiv\; E_\mathrm{loc}(\alpha,\beta,t),
  \label{eq:logderiv}
\end{equation}
where the right-hand side is the standard variational local energy $E_\mathrm{loc}$. We discretize time into a set of points $\{ t_j\}$ and define the residual of Eq.~\eqref{eq:logderiv} at each time
\begin{equation}
  \Delta(\sigma; t_j)
  = i\,\partial_t \log \widetilde{U}_\theta(\sigma; t_j)
    - E_\mathrm{loc}(\sigma; t_j)
\end{equation}
as the objective to be minimized during training.

The time derivative on the left is computed by differentiating the network output with respect to the context tokens $M$ through automatic differentiation, and contracting it with the predicted velocity field $\dot{M}(t_j)$. Indeed, this is where using the FNO to map $H(t)$ to $\dot{M}(t)$ instead of $M(t)$ improves numerical stability, as this choice avoids the discretization dependence associated with numerical differentiation schemes such as finite-difference methods.

We also note that Eq.~\eqref{eq:logderiv} is invariant under a
time-dependent global phase,
${U}_{\alpha\beta}(t)\to
e^{-i\chi(t)} {U}_{\alpha\beta}(t)$. As a consequence,
$i\partial_t\log\widetilde{U}$ in the residual is shifted by the same scalar
$\dot{\chi}(t)$ for every configuration at fixed~$t$.
To quotient out this gauge degree of freedom, we center the
residual by subtracting its sample mean at each point in time. This approach was first explored in Ref.~\cite{tNQS_Carleo}.
The physical loss is then
\begin{equation}
  \mathcal{L}_\mathrm{phys}
  = 
      \bigl|\Delta(t)
            - \bar{\Delta}(t)\bigr|^2,
  \label{eq:phys_loss}
\end{equation}
where
\begin{equation}
  \bar{\Delta}(t)
  = \frac{1}{M}\sum_{m=1}^{M}\Delta(\sigma_m,\,t)
  \label{eq:mean_residual}
\end{equation}
is the mean residual over $M$ samples of spin configurations drawn at time $t$.

The physical loss fixes $\widetilde{U}$ only up to its initial condition.
To enforce $\widetilde{U}(0)=I/\sqrt{D}$, we add an
anchor loss
\begin{equation}
  \mathcal{L}_\mathrm{id}
  = \Bigl\langle
      \bigl|\widetilde{U}_{\alpha\beta}(0)
            - \delta_{\alpha\beta}/\sqrt{D}\,\bigr|^2
    \Bigr\rangle_{\!\alpha,\beta},
  \label{eq:anchor}
\end{equation}
which penalizes deviations from the identity at $t=0$.
In practice, this loss matches diagonal entries
$(\alpha=\beta)$ to $1/\sqrt{D}$ and penalizes off-diagonal entries.
The anchor loss prevents spurious solutions that satisfy the
Schr\"odinger-equation residual but drift from the physical initial condition.
 
The full training objective is therefore
\begin{equation}
  \mathcal{L}_\mathrm{total}
  = \mathcal{L}_\mathrm{phys}
    + \lambda_\mathrm{anchor}\,\mathcal{L}_\mathrm{id},
  \label{eq:total_loss}
\end{equation}
where $\lambda_\mathrm{anchor}$ controls the relative weight
of the initial-condition enforcement.

In practice, we first run a short warm-up phase of training using only the anchor loss $\mathcal{L}_{\mathrm{id}}$
(Eq.~\ref{eq:anchor}), which initializes the model near
the identity propagator at $t = 0$ before the TDVP residual
is introduced. After warm-up, the full loss
$\mathcal{L}_{\mathrm{total}}$ (Eq.~\ref{eq:total_loss}) is used.

During the main stage, training is performed on batches of $B$ driving
trajectories, $K$ time points ($\{t_k\}$) sampled uniformly from the full time
grid, and $M$ doubled-space configurations $\sigma$ that are
autoregressively sampled from the current model at each point in time.  The stochastic estimate of the loss at each optimization step is therefore:
\begin{equation}
  \mathcal{L}
  = \frac{1}{BKM}\sum_{b=1}^{B}\sum_{k=1}^{K}\sum_{m=1}^M
    \Bigl[
      \mathcal{L}_\mathrm{phys}^{(b,k,m)}
      + \lambda_\mathrm{anchor}\,
        \mathcal{L}_\mathrm{id}^{(b,m)}
    \Bigr].
  \label{eq:batch_loss}
\end{equation}
Note that $\mathcal{L}_\mathrm{phys}^{(b,k,m)}$ depends on the sampled time points, driving trajectory, and spin configuration, while the anchor loss
$\mathcal{L}_\mathrm{id}^{(b,m)}$ does not depend on the time index, as it is evaluated only at $t=0$.
 
We emphasize that the training procedure is entirely
self-supervised: no pre-computed data is needed to learn the propagator, and no supervised trajectories associated with particular initial states are used during training. As a consequence, the UNP cannot possibly predict time evolution through memorization;
instead, it must actually learn a propagator that respects the underlying physics in order to make accurate predictions, as
we demonstrate in the next section.

\section{Numerical Results \label{sec:results}}
We consider a time-dependent Hamiltonian $\HH(t)$ acting on spin-$1/2$ degrees of freedom on a square lattice with size $L_x \times L_y$, with open boundary conditions in both directions. Specifically, we focus on the transverse-field Ising model with a longitudinal drive. This model has close connections with experimental platforms such as Rydberg atoms~\cite{TFIM_Rydberg, TFIM_experiment, TFIM_experiment2}. The Hamiltonian is:
\begin{equation}
\HH(t)
=
-J \sum_{\langle i,j\rangle} Z_i Z_j
-h_z(t)\sum_i Z_i
-h_x(t)\sum_i X_i,
\label{eq:hamiltonian}
\end{equation}
where $\langle i,j \rangle$ denotes nearest-neighbor bonds, $X_i (Z_i)$ denote the Pauli-X (Z) operator on site $i$. We fix $J=1$ throughout as the fundamental energy scale; the transverse $h_x(t)$ and longitudinal fields $h_z(t)$ are functions of time.

During training, the driving fields $h_x(t)$ and $h_z(t)$ are sampled at every optimization step from a family of smooth protocols on the interval $t\in[0,T]$. We generate these protocols as random Fourier series. Specifically, the transverse field is parameterized as
\begin{equation}
    h_x(t) = h_{x0} + \sum_{m=1}^{n_{\max}} a_m \sin(m\omega t + \phi_m),
    \label{eq:driving_field}
\end{equation}
and the longitudinal field $h_z(t)$ is sampled analogously. In all experiments, we take $n_{\max}=10$ and set the fundamental frequency scale to $\omega=10J$. The remaining sampling and optimization hyperparameters are provided in Appendix~\ref{app:hyperparameters}. 

\begin{figure*}[ht!]
    \centering
    \includegraphics[width=0.75\linewidth]{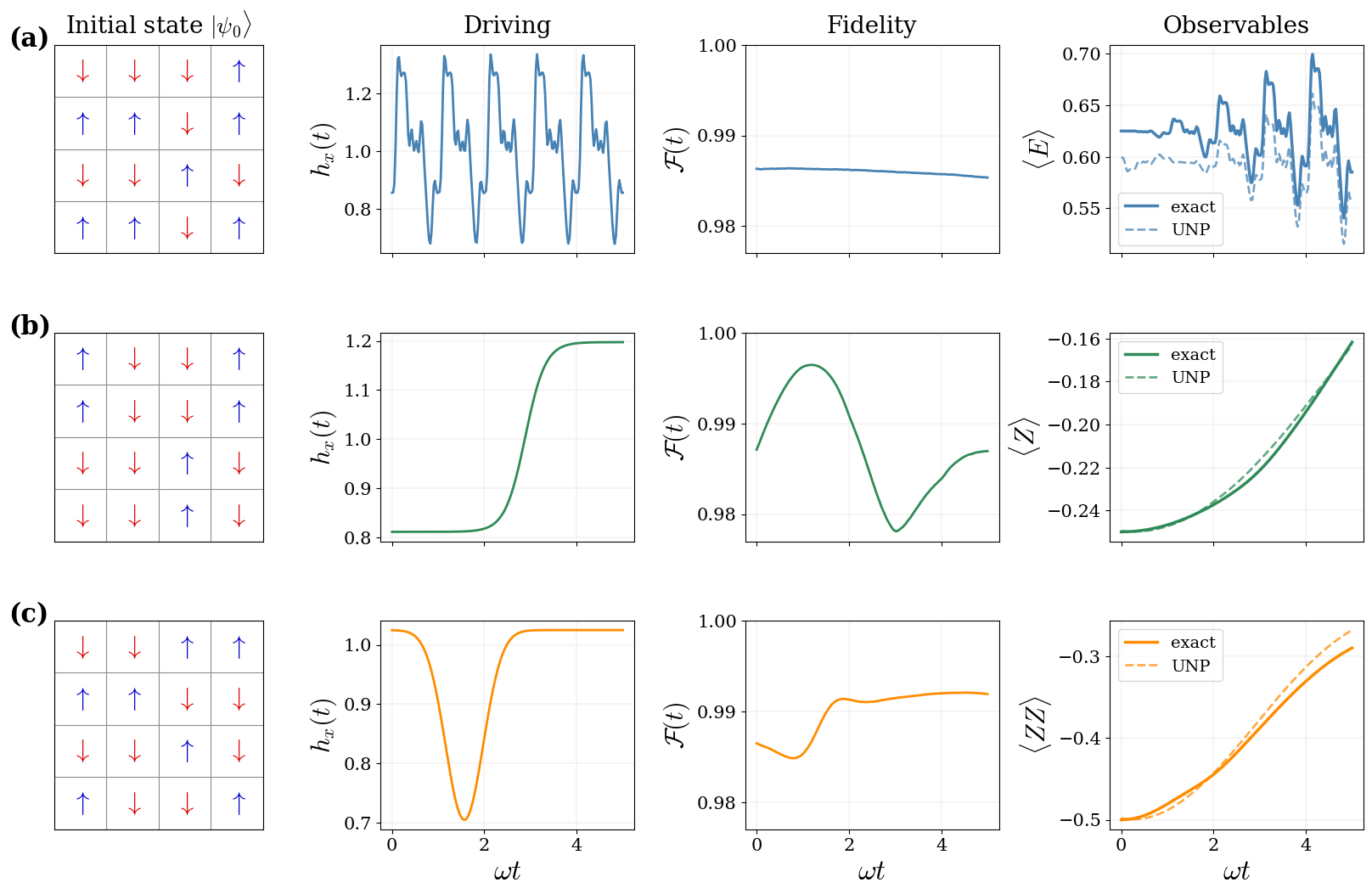}
    \caption{Evaluation of the performance of Universal Neural Propagator for a driven TFIM with system size $4 \times 4$. Each row corresponds to a different computational-basis product state \textit{and} a different time-dependent driving protocol. The four columns show: the initial spin configuration, the driving protocol, the state fidelity \(\mathcal F(t)\) between the exactly evolved state and the UNP-predicted evolution, and a representative local observable, respectively. \textbf{(a)} Performance on an in-distribution driving field. \textbf{(b, c)} UNP predictions for out-of-distribution but experimentally realistic driving protocols: a tanh ramp-up and a Gaussian pulse. The same UNP model is used across all panels without retraining, and demonstrates excellent predictive power across all metrics considered, for all driving protocols, initial states, and at all times throughout the training interval.}
    \label{fig:4x4_results}
\end{figure*}

After training, the UNP model predicts an approximation to the normalized many-body propagator, namely \(\widetilde U(t)\), for any input Hamiltonian $\hat H(t)$. For an initial computational-basis state \(|\beta\rangle\), the predicted, time-evolved wavefunction is obtained from the corresponding column of the learned propagator (Eq.~\ref{eq:psi_from_U_column}). This column-wise interpretation provides a test of whether the learned object behaves as a propagator, rather than as a state-specific model.

We begin with a system of size $4 \times 4$, for which exact time-evolution is possible through direct numerical integration of the Schrödinger equation in the full Hilbert space. We first evaluate the performance of the UNP model on random product states in the computational ($z$) basis. We consider metrics including the state fidelity
\begin{equation}
\mathcal F(t) = \left| \braket{\psi_\text{exact}(t)}{\widetilde{U}(t)|{\psi}_0} \right|^2,
\label{eq:fidelity}
\end{equation}
where $\ket{\psi_0}$ is the initial state, and $\ket{\psi_\text{exact}(t)}$ is the exact time-evolved state.

In addition, we benchmark using expectation values of local observables. In particular, we monitor the spatially averaged magnetization in the $z$- and $x$ direction:
\begin{equation}
\langle Z(t)\rangle
=
\frac{1}{N}\sum_i \langle Z_i (t) \rangle,
\end{equation}
\begin{equation}
    \langle X(t)\rangle
=
\frac{1}{N}\sum_i \langle X_i (t) \rangle,
\end{equation}
as well as the $ZZ$-correlator:
\begin{equation}
\langle ZZ (t)\rangle
=
\frac{1}{N_B}\sum_{\left< i,j \right>} \langle Z_i Z_j (t)\rangle ,
\end{equation}
with $N_B$ the number of bonds and $\left< i, j \right>$ nearest-neighbor pairs. From these local correlators, we can compute the energy $\langle E(t)\rangle$ from Eq.~\ref{eq:hamiltonian}. We note that the metrics we consider are complementary: the fidelity is sensitive to the global quantum state, whereas local observables test whether the learned propagator
reproduces experimentally relevant expectation values.

Figure~\ref{fig:4x4_results} shows representative time evolution starting from three random computational-basis initial states and under three driving protocols. The evolved states $\widetilde U(t) \ket{\psi_0}$ are extracted directly from the corresponding column of the learned propagator; no additional optimization is required. The driving protocol in panel (a) is drawn from the training distribution (Eq.~\ref{eq:driving_field}), while the ramp-up and pulse (modeled by tanh and Gaussian pulses respectively) are \textit{outside} the ensemble used for training. 

Across the three initial states and driving protocols, the UNP prediction remains excellent compared to the exact evolution over the full time interval. The local observables agree almost perfectly, achieving minimal error across the full interval, indicating that the predicted wavefunction has the correct physical correlations. In particular, the model tracks both slowly varying responses and more rapidly oscillating driven dynamics. Furthermore, the fidelity stays consistently above $0.98$ throughout the interval, indicating that the learned doubled-space representation not only reproduces local observables, but also captures the correct phase structure. 

The accuracy of UNP predictions across all three driving protocols demonstrates that the model has learned a functional dependence on the driving history, rather than merely interpolating between a small set of fixed parameters. Furthermore, the transferability across different initial states shows that the learned object behaves as a reusable propagator: the same trained network can be queried for different initial basis states and different driving histories.

\begin{figure}[t]
    \centering
\includegraphics[width=0.75\linewidth]{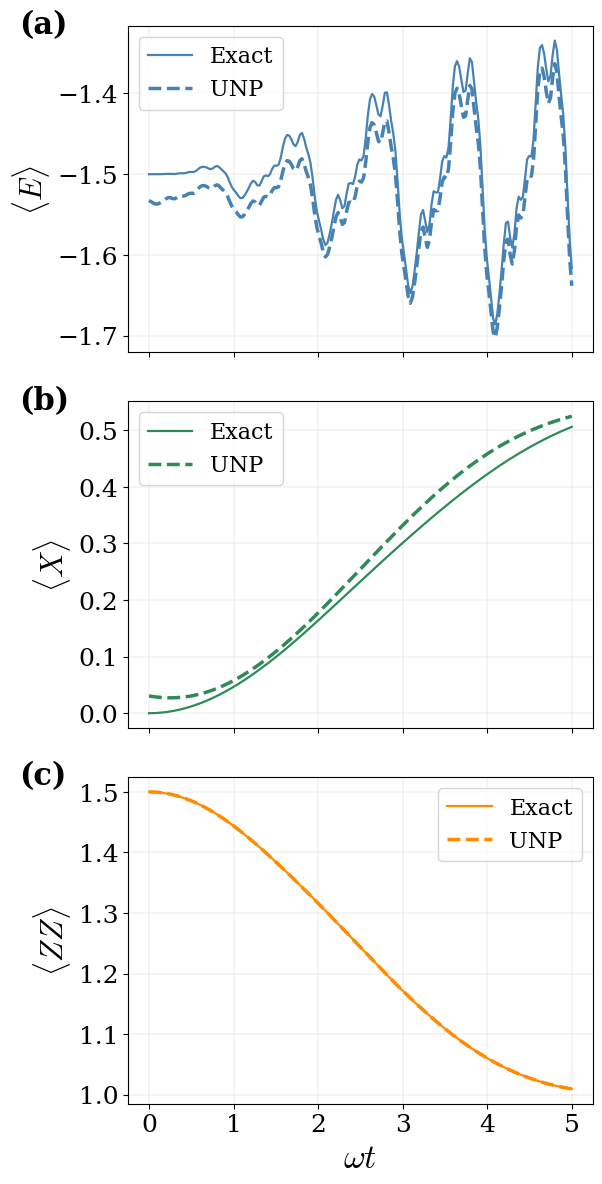}
    \caption{Time-evolution starting from the GHZ state (Eq.~\ref{eq:GHZ}). Because the UNP predicts a propagator, the time evolution of a coherent superposition is obtained by linearly combining the individually propagated computational-basis components. The driving protocols are \textbf{(a)} in the training distribution; \textbf{(b)} a tanh ramp-up; \textbf{(c)} a Gaussian pulse, and are \textit{different} from the ones in Fig.~\ref{fig:4x4_results}. Agreement with exact evolution from this state shows not only the accuracy of individual columns of $\widetilde U(t)$, but also their relative phase structure.}
    \label{fig:GHZ}
\end{figure}

Next, we demonstrate that the propagators predicted by UNP are applicable beyond simple basis states. As a concrete example, we consider the Greenberger-Horne-Zeilinger (GHZ) state~\cite{GHZ}, which is defined as:
\begin{equation}
    \ket{\text{GHZ}} = \frac{1}{\sqrt{2}} \Big( \ket{\uparrow \uparrow \dots \uparrow} + \ket{\downarrow \downarrow \dots \downarrow} \Big),
    \label{eq:GHZ}
\end{equation}
as the initial state. Here $\ket{\uparrow \uparrow \dots \uparrow}$ and $\ket{\downarrow \downarrow \dots \downarrow}$ are ferromagnetically aligned states in the $z$ basis. The GHZ state is entangled and highly nonlocal, as it is a superposition of two macroscopically distinct configurations.
Using the GHZ state as $\ket{\psi_0}$ therefore provides a useful diagnostic of whether the predicted propagator preserves relative phase
information between different columns of the actual propagator.

Despite the entanglement in the initial state, the UNP framework can naturally track its time evolution by exploiting the linearity of the propagator $U(t)$. For a GHZ state, we apply the learned propagator to time-evolve the two ferromagnetic basis states separately, and then superpose the resulting wavefunctions. Once the propagator has been learned, this procedure requires no additional variational optimization for superposed states. This distinguishes UNP from conventional single-state evolution methods: rather than learning a trajectory for one initial state, the model learns a shared representation of $U(t)$, which can be applied to different computational-basis states and combined by linear superposition.

Fig.~\ref{fig:GHZ} shows that
UNP accurately reproduces local observables starting from the GHZ state. Across the three panels, an in-distribution, tanh, and Gaussian driving protocol is used respectively. We note that the driving fields are different from ones shown in Fig.~\ref{fig:4x4_results}. The GHZ benchmark confirms that the learned
doubled-space representation of $U(t)$ captures the relative phase coherence between two macroscopically distinct columns of the propagator, not merely measurement statistics. The accuracy across a variety of driving protocols further reinforces that the UNP successfully learns a functional mapping from $H(t)$ to $U(t)$.

We next demonstrate that the UNP-predicted propagator $\widetilde U(t)$ can be further improved through protocol-specific \textit{fine-tuning}~\cite{finetuning_NQS, Transformer, NOQS}. The central idea is that, once a target driving protocol $H(t)$ is fixed, a small amount of observable data can be used to refine the representation of $\widetilde U(t)$, without retraining the full model. In this way, fine-tuning improves the learned propagator associated with the target protocol, rather than merely fitting the trajectory of a single initial state.

Concretely, we randomly sample initial states from the computational basis. For each state, we compute the corresponding time-dependent observables, namely the  magnetization $\langle X(t)\rangle$ and the correlator $\langle ZZ(t)\rangle$. These values serve as protocol-specific data. During fine-tuning, we optimize only the context tokens $M(t)$, while keeping all parameters in the pretrained FNO and transformer frozen. This approach is motivated by the protocol-specific nature of the task: $H(t)$ enters the transformer entirely through these tokens, making $M(t)$ the primary target for fine-tuning.

The fine-tuning loss function is data-driven:
\begin{align}
    \mathcal{L}_{\mathrm{FT}}
    =
    \frac{1}{M}
    \sum_{m=1}^{M}
    \sum_{t_k}
    \Big[
    &
    \left(
    \langle X_m(t_k) \rangle_{\mathrm{UNP}}
    -
    \langle X_m(t_k) \rangle_{\mathrm{data}}
    \right)^2
    \nonumber \\
    +
    &
    \left(
    \langle ZZ_m(t_k) \rangle_{\mathrm{UNP}}
    -
    \langle ZZ_m(t_k) \rangle_{\mathrm{data}}
    \right)^2
    \Big],
    \label{eq:FT_loss}
\end{align}
where $m$ labels the initial states used for fine-tuning, $t_k$ denotes the discretized time points, and the subscripts ``UNP'' and ``data'' denote model predictions and target observable values, respectively. In an experimental setting, the latter would correspond to measured observables; in our numerical benchmark, they are obtained from exact time evolution. We use $M=20$ randomly sampled initial states.

\begin{figure}
    \centering
    \includegraphics[width=1\linewidth]{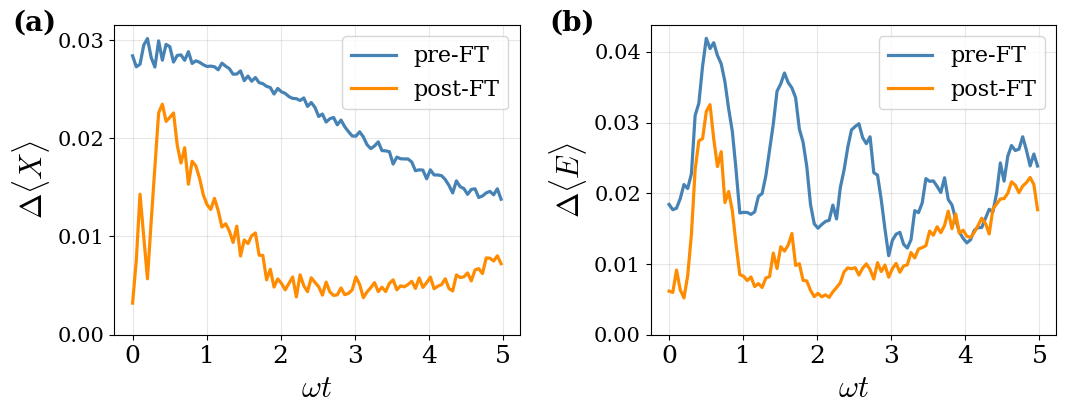}
    \caption{Mean absolute error (MAE) of local observables, before and after fine-tuning the propagator. \textbf{(a)}: MAE of the $x$-magnetization $\langle X \rangle$; \textbf{(b)}: MAE of the energy per spin $\langle  E \rangle$. Both quantities are averaged over 50 random $z$-basis product states that have not been seen during fine-tuning. Fine-tuning using only a small number of initial states consistently improves the propagator's performance across the entire time interval.}
    \label{fig:fine-tuning}
\end{figure}

To test whether the fine-tuned propagator generalizes beyond the measured states, we evaluate the model on $50$ randomly sampled computational-basis states that were not included during fine-tuning. For an observable $\mathcal{O}$, we compare the mean absolute error (MAE)
\begin{equation}
    \Delta \langle \mathcal{O}(t) \rangle
    =
    \frac{1}{N_{\mathrm{test}}}
    \sum_{m=1}^{N_{\mathrm{test}}}
    \left|
    \langle \mathcal{O}_m(t) \rangle_{\mathrm{UNP}}
    -
    \langle \mathcal{O}_m(t) \rangle_{\mathrm{exact}}
    \right|,
    \label{eq:FT_MAE}
\end{equation}
where $N_{\mathrm{test}}=50$, before and after fine-tuning.

Figure~\ref{fig:fine-tuning} shows that fine-tuning the context tokens significantly reduces the MAE for both $\langle X(t)\rangle$ and $\langle E(t)\rangle$ across the full time window. Other observables, such as $\langle Z(t) \rangle$ and $\langle ZZ(t) \rangle$, also achieve reduced error after the data-driven tuning stage, as reflected by the reduction in MAE for $\langle E(t) \rangle$. Importantly, this improvement is measured on \textit{unseen} initial states, demonstrating that the fine-tuning procedure does not simply memorize the supervised trajectories. Instead, the observable data from only $20$ basis states is sufficient to improve the protocol-conditioned propagator itself. Since the full Hilbert space contains $2^{16} = 65536$ basis states, our result suggests that very sparse protocol-specific data can be sufficient to tune the learned propagator in a highly data-efficient manner.

Finally, we demonstrate the scalability of our UNP framework by studying a system of size $6\times 6$. The Hilbert space is too large for exact methods, so we benchmark the predictions of local observables against time-dependent Density Matrix Renormalization Group (tDMRG). Details about tDMRG implementation are in Appendix~\ref{app:tDMRG}.

As shown in Fig.~\ref{fig:6by6}, the UNP remains accurate for driving protocols that are both within and outside of the training distribution. The predicted local magnetizations and nearest-neighbor correlators
match the tensor-network results closely over the time window. This provides evidence that our framework is not restricted to small Hilbert spaces where the full propagator can be
explicitly computed. Instead, the UNP can be scaled up to make accurate predictions for larger system sizes, in spite of the exponentially larger state space.

We briefly comment on computational cost. All UNP models are trained on a single NVIDIA H100 NVL GPU (96 GB VRAM). Training takes approximately 4.8 hours for the $4 \times 4$ system and 16.5 hours for the $6\times 6$ system. By contrast, a single tDMRG run for one initial state and driving protocol on the $6 \times 6$ system takes approximately 3 hours. Crucially, the UNP training cost is a one-time expense: once trained, the model can be queried for any combination of driving protocol and initial state at negligible additional cost. For example, evaluating local observables takes a runtime on the order of seconds, whereas tDMRG must be re-performed from scratch for each new configuration.

We stress again that \textit{no} time-evolved physical states are used as supervised targets during training. Therefore, the remarkable generalization across diverse protocols and initial conditions does not come from the network memorizing specific state dynamics, but rather from a genuine learning of the underlying map between functional spaces of driving protocols and propagators.

\begin{figure}[t]
    \centering
    \includegraphics[width=1\linewidth]{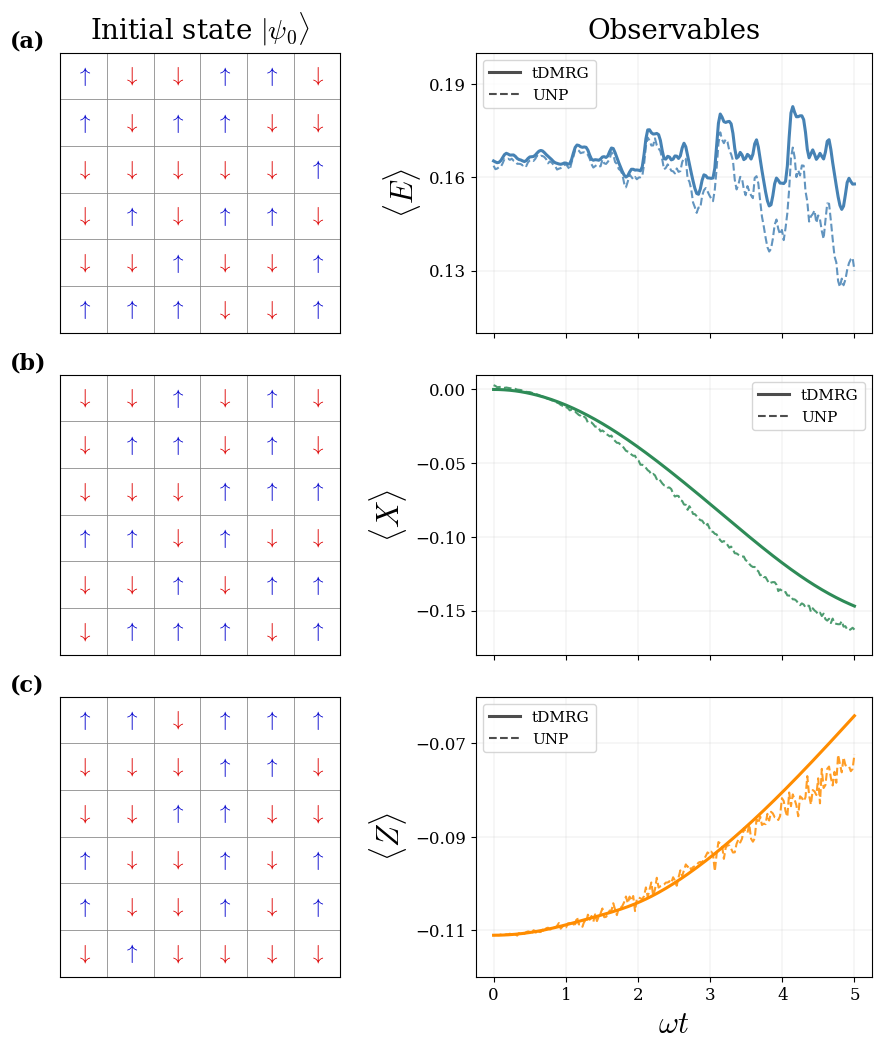}
    \caption{UNP performance on a larger system size ($6\times 6$) that is beyond the reach of exact methods. The driving protocols across the three panels are the same as those in Fig.~\ref{fig:4x4_results}. Despite the exponentially larger space of basis states, the UNP model still successfully captures local observables under time evolution for randomly generated initial states, when compared against tensor-network-based methods.}
    \label{fig:6by6}
\end{figure}

\section{Discussion and Outlook \label{sec:discussion}}
In this work, we introduce the Universal Neural Propagator, a novel framework that represents many-body quantum time evolution directly, rather than treating single time-evolved states. The key is to represent the normalized propagator as a neural quantum state in a doubled Hilbert space, and learn the functional mapping from driving protocols $H(t)$ to propagators $U(t)$.

We validate the UNP framework on a two-dimensional driven Ising model. For \(4\times4\) systems, a single trained model accurately reproduces dynamics for multiple computational-basis initial states and for both in-distribution and out-of-distribution driving protocols. The predicted states reproduce local observables and maintain high fidelity with exact evolution. We also show that sparse protocol-specific observable data can be used to fine-tune the context trajectory and improve predictions on unseen initial states. Finally, for a \(6\times6\) system beyond exact diagonalization, comparison with tDMRG shows that the model continues to capture local dynamical observables. 

Together, these results establish UNP as a concrete step toward transferable, neural-network based simulation of many-body dynamics: after self-supervised training, the same model can be reused across driving protocols and initial states within the trained problem class, without solving a new variational time-evolution problem. Our framework is especially promising in settings where many dynamical queries are required, such as driven quantum simulation, protocol design, and quantum control.

Despite the wide applicability of the propagator across an exponentially large set of initial states, it can in fact be represented at a rather modest cost. Indeed, representing a propagator in doubled space has the same formal cost as representing a density matrix in Liouville space: for $N$ spin-1/2 degrees of freedom, each doubled-space token encodes a pair of physical indices, so the local vocabulary increases from 2 to 4, while the sequence length remains $N$~\cite{liouville_space}. Equivalently, this can be viewed as a patched transformer acting on a system of size $2N$, where each patch groups two sites into a single four-level token~\cite{patched_transformer}. Thus, the propagator does not require a qualitatively new scaling beyond standard neural representations of mixed states or vectorized operators.

While the UNP in principle allows for evolving arbitrary initial states by linearity, in practice, this linear reconstruction can become expensive for dense initial states. For example, an \(x\)-polarized product state is a uniform superposition of all \(2^N\) computational-basis states in the \(z\) basis. Directly evaluating the evolution of such a state would require an exponentially large sum and therefore become infeasible. This is not a failure of the learned propagator; however, it does highlight the importance of choosing a convenient basis. In practice, one can choose an appropriate basis to train the UNP model to better suit the desired initial state.

There are a few natural ways to extend our work. One direction is data-assisted refinement of the \textit{entire} UNP model. In Sec.~\ref{sec:results}, we showed that for a fixed protocol, optimizing only the context trajectory \(M(t)\) using a small number of observable trajectories improves predictions on unseen initial states. A natural extension is to use experimental or numerical data collected across \textit{multiple} driving protocols \textit{and} initial states to fine-tune the weights in FNO and transformer themselves. Such a tuning procedure could convert measurements into a globally improved universal neural propagator and potentially enhance generalization to \textit{both} unseen protocols and unseen initial conditions.

Another promising direction is to extend the UNP framework to open quantum systems. Under Lindblad dynamics, the object of interest is a quantum channel mapping \(\rho(0)\) to \(\rho(t)\), rather than a unitary propagator mapping \(|\psi(0)\rangle\) to \(|\psi(t)\rangle\). Such channels can be vectorized as superoperators acting in Liouville space, suggesting a natural extension analogous to the construction used in this work. A neural representation of these superoperators could provide a route toward foundation models for driven dissipative dynamics, transferable across both time-dependent controls and initial density matrices.

More broadly, the success of UNP in our work suggests that neural quantum simulation can be generalized beyond individual state trajectories and Hamiltonians. By learning dynamical maps directly, machine-learning architectures may provide a scalable route toward simulation, calibration, and control of complex quantum systems.

\begin{acknowledgments}
YP is supported by the US National Science Foundation (NSF) Grants  No.\ PHY-2216774 and No.\ DMR-2406524.
\end{acknowledgments}

\appendix
\section{Architectural Details \label{app:architecture_details}}
As discussed in Sec.~\ref{sec:architecture}, the UNP architecture consists of a Fourier Neural Operator (FNO) and transformer. In this Appendix, we discuss details about the inner workings of both architectures.

We begin with the FNO, which processes driving protocols $H(t)$ and distills them into a set of context tokens $M(t)$. Neural operators are generalizations of neural networks: instead of mappings between finite-dimensional vector spaces, neural operators map between infinite-dimensional function spaces acting on a domain $D$~\cite{NeuralOperator_general, NeuralOperator2}. In this work, the domain is restricted to be $D=(0,T)$, i.e. inputs and outputs are both functions of time. In particular, the neural operator in the UNP architecture maps from the time-dependent Hamiltonian $H(t)$ to context tokens $\dot M(t)$:
\begin{equation}
    \mathcal{A}: H(t) \rightarrow \dot M(t).
\end{equation}

The context tokens $M(t)$ are obtained from the velocity field
$\dot{M}(t)$ via numerical integration of
Eq.~(\ref{eq:token_integration}) using Simpson's method. In practice, all functions such as $H(t)$ are evaluated on a uniform grid of $N_T$ time points.

\begin{figure}
    \centering
\includegraphics[width=1.0\linewidth]{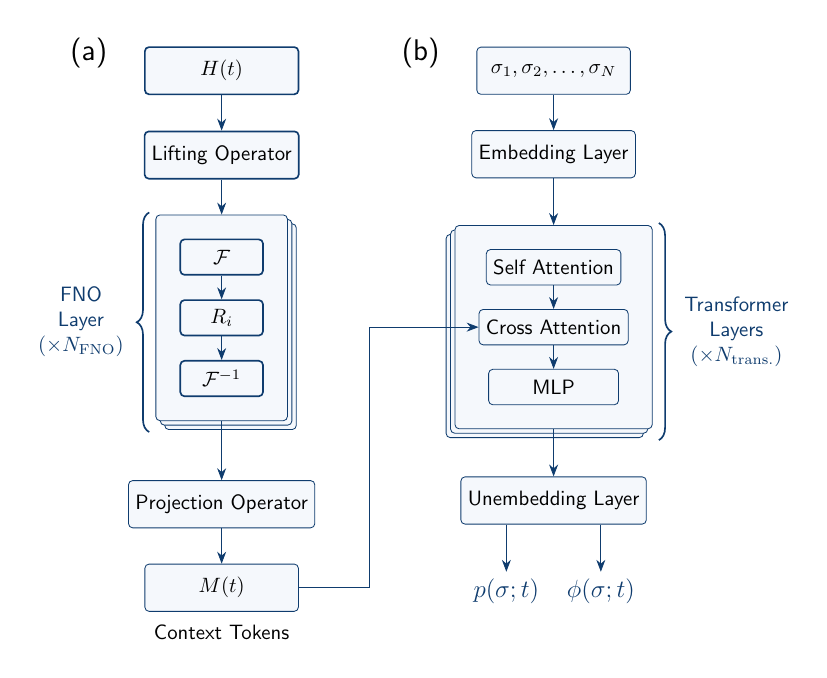}
    \caption{Illustration of the architectural detail of the UNP model. \textbf{(a)}~The FNO maps driving protocols $H(t)$ to the context token
velocity $\dot{M}(t)$, which is integrated to obtain $M(t)$. The input
is lifted to a latent space, passed through $N_\text{FNO}$ layers that apply learnable spectral convolution, and projected to the output function.
\textbf{(b)}~The transformer processes doubled-space token sequences
$\sigma_1,\sigma_2,\dots,\sigma_N$ through an embedding layer, followed
by $N_\text{trans.}$ decoder blocks, each consisting of masked
self-attention, cross-attention to the context tokens $M(t)$, and a
feed-forward network. The unembedding layer produces the conditional
amplitude $p(\vec\sigma;t)$ and phase $\phi(\vec\sigma;t)$, which are
combined to yield the propagator matrix elements.}
    \label{fig:architectural_details}
\end{figure}

The architecture of FNO consists of three parts: a lifting operator, FNO layers, and projection operator, as shown in Fig~\ref{fig:architectural_details}(a). First, the input function $H(t)$ is concatenated with a normalized time coordinate $t/T \in [0,1]$ and mapped to the latent space with dimension $d_\text{FNO}$, through a pointwise lifting operator $L$. The same operator is applied identically to each point in time. Denoting the dimension of input function $H(t)$ as $d_\text{in}$ and the number of time points $N_t$, $L$ is then a tensor of shape $(d_\text{in} +1, d_\text{FNO})$. The output of the lifting operation is a function $u_0(t)$, the first hidden state.

Next, the hidden states are iteratively updated by $N_\text{FNO}$ layers. Each layer is parameterized in frequency domain and acts on the input function $u_i(t)$ as:
\begin{equation}
    u_{i+1}(t) = \sigma \left( \mathcal{F}^{-1}(R_{i} \cdot \mathcal{F}(u_i)) (t) + b_i(t) \cdot u_i(t) \right),
    \label{eq:fourier_layer}
\end{equation}
where $R_i$ is the Kernel function parameterized in frequency domain, $b_i$ is a pointwise linear transformation, and $\mathcal{F}$ and $\mathcal{F}^{-1}$ denote the Fourier transform and the inverse Fourier transform, respectively. In practice, we implement $\mathcal{F}$ and $\mathcal{F}^{-1}$ using (inverse) Fast Fourier Transforms (FFT), which speed up computations significantly. After the Fourier transform $\mathcal{F}$, we also apply spectral filtering and truncate the number of frequency modes. We keep $k_\text{max}$ modes with the lowest frequencies before passing to $R_i$, which is a tensor of shape $(k_\text{max}, d_\text{FNO}, d_\text{FNO})$. Crucially, information from higher frequency modes can still be recovered due to the nonlinearities present in the model~\cite{FNO_Li}.

Finally, a pointwise projection operator
$P$ of shape $\mathbb{R}^{d_\text{FNO} \times (N_\text{context} \cdot d_e)}$ is
applied uniformly at each time point, mapping the last hidden state to the output function.
The raw output at each time $t$ is a vector of dimension
$N_\text{context} \cdot d_e$; it is then reshaped into $N_\text{context}$
context tokens, each of dimension $d_e$. This ensures dimensional compatibility with the transformer's cross-attention
mechanism, which queries $M(t)$ using $d_e$-dimensional representations.

Next we discuss the Transformer part of the UNP architecture. The transformer is an architecture that is based on the self-attention mechanism. It has attracted much attention in recent years due to applications to natural language processing~\cite{vaswani2017attention}. Transformer-based wavefunction ansatz has also proven remarkably expressive~\cite{transformer_wf_2, transformer_wf_3, Transformer, patched_transformer}.

As shown in Fig.~\ref{fig:architectural_details}(b), the transformer also consists of three parts. The inputs are sequences of doubled-space tokens $\sigma_1, \sigma_2, \dots, \sigma_N$, where each $\sigma_i \in \{0,1,2,3\}$. Here we adopt a snake ordering, which traverses the two-dimensional
lattice in a zigzag pattern, ensuring that
consecutive sites in the sequence are spatially adjacent. The ordered tokens first go through an embedding layer. Each local degree of freedom is converted into a
\(d_e\)-dimensional vector representation by selecting the corresponding column of a shared embedding matrix,
\begin{equation}
    \mathbf{e}_i = W_E [\sigma_i], \qquad W_E \in \mathbb{R}^{d_e \times 4}.
    \label{eq:embedding}
\end{equation}
Here \(W_E\) denotes a trainable embedding parameter that is applied uniformly at
every lattice site.

Because the same embedding map is used at all sites, the spin value alone does
not specify where the spin is located in the lattice. However, spatial locality is important for encoding the underlying physical structure; therefore, we add a learnable positional vector
\(\mathbf{p}_i\in\mathbb{R}^{d_e}\) to each embedded spin. The input token at site
\(i\) is therefore
\begin{equation}
    \mathbf{x}_i = \mathbf{e}_i + \mathbf{p}_i,
    \qquad i=1,\ldots,N .
    \label{eq:pos_encoding}
\end{equation}
Collecting these row vectors gives the transformer input
\(X\in\mathbb{R}^{N\times d_e}\).

The sequence of token representations is then passed through \(N_\text{trans.}\) decoder
blocks. Each block contains three main operations: masked multi-head
self-attention, cross-attention to context tokens produced by the FNO, and a
position-wise feed-forward network. Each operation is equipped with a residual connection and layer normalization. In the following, we describe one representative
decoder block and suppress the block index for notational clarity.

The first operation is masked multi-head self-attention. For each attention head
\(s=1,\ldots,n_h\), the input matrix \(X\) is linearly projected into queries,
keys, and values:
\begin{equation}
    Q^{(s)} = X W_Q^{(s)}, \qquad
    K^{(s)} = X W_K^{(s)}, \qquad
    V^{(s)} = X W_V^{(s)} ,
    \label{eq:self_attn_proj}
\end{equation}
where
\(W_Q^{(s)}, W_K^{(s)}, W_V^{(s)}\in\mathbb{R}^{d_e\times d_h}\), with
\(d_h=d_e/n_h\) the feature dimension assigned to each head. The attention map
for the \(s\)-th head is computed as
\begin{equation}
    \mathrm{Attn}^{(s)}
    =
    \mathrm{softmax}\!\left(
        \frac{Q^{(s)} {K^{(s)}}^T}{\sqrt{d_h}} + \mathcal{M}
    \right)V^{(s)} ,
    \label{eq:self_attn}
\end{equation}
with the softmax applied independently to each row. For a vector
\(A\in\mathbb{R}^d\), the softmax is defined as
\begin{equation}
    \mathrm{softmax}(A)_i
    =
    \frac{\exp(A_i)}{\sum_{k=1}^{d}\exp(A_k)} ,
    \label{eq:softmax}
\end{equation}
so that the entries are transformed into normalized positive weights.

The causal mask
\(\mathcal{M}\in\mathbb{R}^{N\times N}\) is defined by
\begin{equation}
    \mathcal{M}_{ij}
    =
    \begin{cases}
        0, & j \leq i, \\
        -\infty, & j > i .
    \end{cases}
    \label{eq:causal_mask}
\end{equation}
This mask prevents token \(i\) from accessing information from later sites in
the ordering. Consequently, the representation associated with \(\sigma_i\)
depends only on the preceding spins \(\sigma_1,\ldots,\sigma_{i-1}\), thereby
maintaining the conditional factorization in Eq.~\eqref{eq:ar_prob}. Such
autoregressive masking is essential for normalized sequential representations of
many-body configurations~\cite{autoregressive-hubbard, luodi_autoregressive_open,
krylov_transformer, sharir2020_autoregressive}.

The outputs from all attention heads are concatenated and projected back to the
model dimension:
\begin{equation}
    \mathrm{SelfAttn}(X)
    =
    \mathrm{Concat}\!\left(
        \mathrm{Attn}^{(1)},\ldots,\mathrm{Attn}^{(n_h)}
    \right) W_O + b_O ,
    \label{eq:self_attn_concat}
\end{equation}
where \(W_O\in\mathbb{R}^{d_e\times d_e}\) and
\(b_O\in\mathbb{R}^{d_e}\). The resulting output is combined with the block input
through a residual connection and then normalized.

The second operation in the decoder block is cross-attention. This layer injects
time-dependent information into the spin sequence by attending to temporal
features generated by the Fourier Neural Operator. Concretely, cross-attention is implemented as the second sub-layer of each decoder block, following self-attention. Let
$X' \in \mathbb{R}^{N \times d_e}$ denote the output of the self-attention sub-layer, and let
$M(t) \in \mathbb{R}^{N_c \times d_e}$ denote the context matrix produced by the neural operator $\mathcal{N}$ at time $t$.
In this sub-layer, the transformer representations act as queries, while the context tokens provide the keys and values.
Thus, for each of the $n_h$ attention heads, we define
\begin{equation}
    Q^{(c)} = X' W_Q^{(c)}, \quad
    K^{(c)} = M(t) W_K^{(c)}, \quad
    V^{(c)} = M(t) W_V^{(c)},
    \label{eq:cross_attn_proj}
\end{equation}
where
$W_Q^{(c)}, W_K^{(c)}, W_V^{(c)} \in \mathbb{R}^{d_e \times d_h}$.
The superscript $(c)$ indicates that these parameters belong to the cross-attention mechanism.

The resulting cross-attention output for each head is
\begin{equation}
    \mathrm{Attn}^{(c)}
    =
    \mathrm{softmax}\!\left(
    \frac{Q^{(c)} {K^{(c)}}^\top}{\sqrt{d_h}}
    \right) V^{(c)}.
    \label{eq:cross_attn}
\end{equation}
The attention weights have shape $\mathbb{R}^{N \times N_c}$, so each of the $N$ spin tokens attends over all $N_c$ context tokens.
Unlike in the causal self-attention sub-layer, no causal mask is applied here: every spin token is allowed to access the full temporal context encoded in $M(t)$.
Finally, the outputs from all heads are concatenated and linearly projected, followed by a residual connection and layer normalization, as in the self-attention sub-layer; see Eq.~\ref{eq:self_attn_concat}.

The third operation is a position-wise feed-forward network,
\begin{equation}
    \mathrm{FFN}(\mathbf{x})
    =
    W_2\,\sigma(W_1\mathbf{x}+\mathbf{b}_1)+\mathbf{b}_2 ,
    \label{eq:ffn}
\end{equation}
where \(W_1\in\mathbb{R}^{d_f\times d_e}\),
\(W_2\in\mathbb{R}^{d_e\times d_f}\), \(d_f\) is the hidden width of the
feed-forward layer, and \(\sigma(\cdot)\) denotes the GeLU activation. As in the
attention mechanism, this sub-layer is followed by a residual connection and layer
normalization.

Finally, the latent space state after the last decoder block is mapped to conditional log-probabilities $\log p(\sigma_i|\sigma_{<i};\,t)$ 
and conditional phases $\phi(\sigma_i|\sigma_{<i};\,t)$ by two separate linear heads, applied uniformly to all sites, each of shape $\mathbb{R}^{4 \times d_e}$. A softmax is applied to the logits to obtain the conditional probabilities. The amplitude and phase are combined to produce the matrix element $U_{\alpha \beta}(t)$.

\section{Hyperparameters \label{app:hyperparameters}}
To ensure reproducibility, we list the hyperparameters used in this work in this appendix. The choices of hyperparameters in the model, as well as for pre-training and fine-tuning, are included in Table~\ref{tab:parameters}.

\begin{table}[ht]
\centering
\begin{tabular}{ll}
\hline
\textbf{} & \textbf{Hyperparameter} \\ \hline
\textbf{Architecture (Transformer)} & \\
Number of Decoder Layers ($N_{\text{trans.}}$) & 3 \\
Embedding Dimension ($d_e$) & 96 \\
Number of Attention Heads ($n_h$) & 8 \\
Feed-forward Dimension ($d_f$) & $4 \times d_e = 384$ \\
Activation Function & GeLU~\cite{gelu} \\
\hline
\textbf{Architecture (FNO)} & \\
Number of FNO Layers ($N_\text{FNO}$) & 3 \\
FNO Width ($d_\text{FNO}$) & 128 \\
FNO frequency modes ($k_\text{max}$) & 48 \\
Number of Context Tokens ($N_C$) & 4 \\
Number of Points in Time & 200\\
\hline
\textbf{Training} & \\
Optimizer & Adam~\cite{adam} \\
Initial Learning Rate (LR) & $5 \times 10^{-4}$ \\
LR decay factor & 0.95 \\
LR decay rate & 2000 (steps) \\
Minimum LR & $5 \times 10^{-6}$ \\
Batch size ($B$) & 6 \\
Time points per step $(K)$ & 4 \\
MC sample per step $(M)$ & 128 \\
Training Steps & 120,000 \\
Weight of Anchor Loss ($\lambda_\text{anchor}$) & 1.0 \\
Gradient Clipping & 0.1 \\
\hline
\textbf{Fine Tuning} & \\
Training Steps & 500 \\
Learning Rate & $1 \times 10^{-3}$ \\
\hline
\end{tabular}
\caption{\label{tab:parameters}Hyperparameters for the UNP model, for system size $4 \times 4$. For the system size $6 \times 6$, the only changes are in the number of decoders, FNO layers, context tokens, and: $N_\text{trans.}=3\rightarrow 4$; $N_\text{FNO} = 3\rightarrow 4$; $N_C =4 \rightarrow 8$. The batch size $B$ and sampled time points $K$ are reduced to $4$ and $3$, respectively, to fit within VRAM limits.}
\end{table}

The driving protocols $h_x(t)$ and $h_z(t)$ are sampled independently
at each training step. For the transverse field,
\begin{equation}
  h_x(t) = h_{0}^{(x)}
    + \sum_{m=1}^{n_{\max}} a_m^{(x)}\, m^{-3/2}\,
      \sin\bigl(m\,\omega_0\, t + \phi_m^{(x)}\bigr),
  \label{eq:hx_detail}
\end{equation}
where $n_{\max} = 10$ and $\omega_0 = 10J$.
The offset $h_{0}^{(x)}$ is drawn uniformly from $[0.95,\,1.05]\,J$,
the amplitudes $a_m^{(x)}$ are drawn uniformly from $[-0.50,\,0.50]\,J$,
and the phases $\phi_m^{(x)}$ are uniform on $[0, 2\pi)$.
The longitudinal field $h_z(t)$ is generated analogously, but with
a smaller amplitude range $[-0.05,\,0.05]\,J$ and offset $h_{0}^{(z)}$
drawn uniformly from $[-0.05,\,0.05]\,J$.

The entire training procedure is done on a single NVIDIA H100 NVL with 96 GB VRAM. The training time for the $4\times 4$ system is 4.8 hours. For the $6 \times 6$ system, due to the larger lattice and more parameters in the UNP, the run time is approximately 16.5 hours.

\section{Implementation of tDMRG \label{app:tDMRG}}

For the $6 \times 6$ system (Fig.~\ref{fig:6by6}), exact
diagonalization is infeasible. We benchmark the UNP predictions against time-dependent density matrix renormalization group (tDMRG), a tensor-network method that represents the quantum state as a matrix product state (MPS) and evolves it via a Suzuki-Trotter decomposition
of the time-evolution operator~\cite{dmrg_review1, dmrg1, dmrg2}.
 
We implement tDMRG using the TeNPY package~\cite{tenpy, tenpy2024}. The 2D lattice is mapped to a one-dimensional MPS using the same snake ordering as the transformer input. The bond dimension is chosen to be $\chi = 128$, and we have verified that the results converge in $\chi$.
 
The implementation of tDMRG is performed using 64 threads on an AMD EPYC 9454P server. Evolving an initial product state under any driving protocol takes a machine runtime of 165.7 minutes.

\bibliography{main}

\end{document}